\documentclass[12pt]{article}

\usepackage{amssymb,amsmath,amsthm}
\usepackage{natbib}
\usepackage[dvips]{graphicx}
\usepackage[margin=0.5cm,font={footnotesize,sf},labelfont={footnotesize,sf,sc}]{caption}
\DeclareMathOperator*{\argmin}{argmin}

\begin{document}


\title{\vspace*{-20mm} Investigations of a compartmental model for leucine kinetics using nonlinear mixed effects models with ordinary and stochastic differential equations}

\author{Martin Berglund\thanks{\scriptsize{Mathematical Sciences, University of Gothenburg, Gothenburg, Sweden}}~\thanks{\scriptsize{Mathematical Sciences, Chalmers University of Technology, Gothenburg, Sweden}} \and Mikael Sunn\aa ker\thanks{\scriptsize{Fraunhofer-Chalmers Research Centre for Industrial Mathematics, Gothenburg, Sweden}}~\thanks{\scriptsize{Department of Biosystems Science and Engineering, ETH Zurich, Switzerland (current address)}}~\thanks{\scriptsize{Competence Center for Systems Physiology and Metabolic Diseases, Zurich, Switzerland (current address)}} \and Martin Adiels\thanks{\scriptsize{Wallenberg Laboratory for Cardiovascular Research, University of Gothenburg, Gothenburg, Sweden}} \and  Mats Jirstrand$^{\ddagger}$ \and Bernt Wennberg$^{*\dagger}$}

\maketitle

\begin{abstract}

\noindent
Nonlinear mixed effects models represent a powerful tool to simultaneously analyze data from several individuals. In this study a compartmental model of leucine kinetics is examined and extended with a stochastic differential equation to model non-steady state concentrations of free leucine in the plasma. Data obtained from tracer/tracee experiments for a group of healthy control individuals and a group of individuals suffering from diabetes mellitus type~$2$ are analyzed. We find that the interindividual variation of the model parameters is much smaller for the nonlinear mixed effects models, compared to traditional estimates obtained from each individual separately. Using the mixed effects approach, the population parameters are estimated well also when only half of the data are used for each individual. For a typical individual the amount of free leucine is predicted to vary with a standard deviation of~$8.9\%$ around a mean value during the experiment. Moreover, leucine degradation and protein uptake of leucine is smaller, proteolysis larger, and the amount of free leucine in the body is much larger for the diabetic individuals than the control individuals. In conclusion nonlinear mixed effects models offers improved estimates for model parameters in complex models based on tracer/tracee data and may be a suitable tool to reduce data sampling in clinical studies.\\\\


\end{abstract}

\section{Introduction} \label{Sec:Introduction}
The powerful combination of nonlinear mixed effects (NLME) models and stochastic differential equations (SDEs) has lately received increasing attention \citep{Donnet:2010,Picchini:2010}, with many applications in the area of pharmacokinetic/pharmacodynamic modelling \citep{Overgaard:2005,Moller:2010}.

Nonlinear mixed effects models are used to study population characteristics when data have been gathered for multiple individuals governed by the same intraindividual (within individuals) mechanisms \citep[see, e.g.,][]{Davidian:1995,Davidian:2003,Olofsen:2004,Pillai:2005,Pinheiro:2000,Sheiner&Beal:1980}. NLME models are hierarchical in structure and take population behaviour and individual properties into account simultaneously. They provide a statistical framework to separate intra- and interindividual (between individuals) variability by using probability distributions for the individual-specific parameters. This stands in contrast to what \cite{Sheiner&Beal:1980} call the two stage approach to population modelling, where model parameters are estimated separately for each individual.

Therefore NLME models give better estimates of the variabilities and it has been shown that the interindividual variabilities of the parameters are better estimated for NLME models than for two stage models \citep[see, e.g.,][]{Olofsen:2004,Sheiner&Beal:1980,Sheiner&Beal:1983,Steimer:1984}. Moreover, as long as the number of individuals is sufficiently large, a smaller amount of data samples is needed for each individual in order to obtain good estimates of population properties. This was, e.g., shown by \cite{Overgaard:2005} in a simulation study of a one-compartment pharmacokinetics model based on a single SDE. Furthermore, \cite{Jonsson:2000} show that NLME models perform better than two stage models to detect and characterize nonlinearities in the physiological system.

NLME models are traditionally based on a function that is nonlinear in the population (fixed effects) parameters and in the individual-specific parameters \citep{Davidian:1995}. A term is also added to accommodate for noise in the measurements. Ordinary differential equations (ODEs) are often used to represent state variables in dynamic models. The model output may be a function of the state, time, model parameters, and a measurement noise term. To accommodate for noise that is intrinsic to the system, and not only due to measurement errors, a better approach may be to use stochastic differential equations. Such intrinsic noise may be due to true randomness in the model parameters and state variables over time. Also in systems without intrinsic noise, stochastic differential equations can be used to reduce model errors due to an oversimplified ODE model.

\begin{figure}[b!]
\setlength{\unitlength}{1mm}
\begin{picture}(140,30)
 \put(33.5,15){\circle{13}}
 \put(32.5,14){$2$}
 \put(33.5,8){\vector(0,-1){8}}
 \put(34.5,4){\scriptsize $k_{L2}$}
 \put(40.5,18){\vector(1,0){14}}
 \put(45,19){\scriptsize $k_{12}$}
 \put(54,12){\vector(-1,0){14}}
 \put(45,13){\scriptsize $k_{21}$}

 \put(61.5,15){\circle{13}}
 \put(60.5,14){$1$}
 \put(46.5,30){\vector(1,-1){10}}
 \put(51,26){\scriptsize $U_1$}
 \put(66.5,20){\vector(1,1){10}}
 \put(67.5,26){\scriptsize $k_{01}$}
 \put(68.5,18){\vector(1,0){14}}
 \put(73,19){\scriptsize $k_{31}$}
 \put(82.5,12){\vector(-1,0){14}}
 \put(73,13){\scriptsize $k_{13}$}

 \put(89.5,15){\circle{13}}
 \put(88.5,14){$3$}
 \put(96.5,18){\vector(1,0){14}}
 \put(101,19){\scriptsize $k_{43}$}
 \put(110.5,12){\vector(-1,0){14}}
 \put(101,13){\scriptsize $k_{34}$}

 \put(117.5,15){\circle{13}}
 \put(116.5,14){$4$}
\end{picture}
\caption{{\footnotesize A graphical representation of the leucine kinetics model. It was developed by \cite{Demant:1996}, based on the work in \cite{Cobelli:1991}. Compartment $1$ corresponds to the amount of free leucine in plasma. Compartments $3$ and $4$ correspond to a body protein pool with leucine uptake and slow release back to the plasma. Compartment $2$ is a hepatic intracellular compartment from which the liver is fed with leucine that is used when apolipoproteins are synthesized. The fractional transfer coefficient between compartments $j$ and $i$ is denoted $k_{ij}$, and $U_1$ is the continuous inflow of material into the first compartment. A more detailed description of the model is given in Section~\ref{Sec:leucine model}.}}
\label{fig:leuModel}
\end{figure}
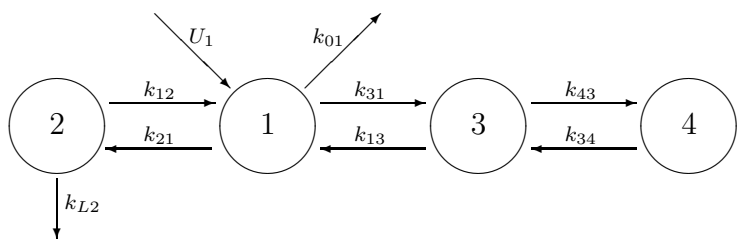

In this study compartmental models and tracer/tracee experiments are analyzed using techniques from nonlinear mixed effects modelling. As an example we focus on a model of leucine kinetics based on tracer/tracee data. Figure~\ref{fig:leuModel} shows a graphical representation of the model. In several publications this model, or minor modifications thereof, has been used as a part of larger models, in which the kinetics of lipoproteins in human blood plasma has been studied \citep[see, e.g.,][for a review]{Barrett:2006}. However, to our knowledge, it is the first time the model is used in an NLME setting.

The evolution over time of the model state variables is described by a system of four coupled ODEs. In this paper we also extend the ODE model with a mean-reverting Ornstein-Uhlenbeck process to model a non-steady state concentration of free leucine in the plasma through the experiment, which turns the system of ODEs into a system of stochastic differential equations, SDEs. As far as we know, it is the first time this approach is used on tracer/tracee data. Moreover we believe that there is no published material, up to date, where NLME and SDEs are combined in such a complex model (w.r.t. the number of model state variables and unknown parameters to estimate).

Experimental data from $19$ control individuals and $15$ individuals suffering from diabetes mellitus type $2$ (DM2) are used to estimate the parameters. Furthermore we investigate if the NLME parameters can be estimated when only a subset of the data are used for each individual. The data have earlier been analyzed in \cite{Adiels:2005_1,Adiels:2005_2,Adiels:2006} in a larger ODE based compartmental model.

We are aware of only a few studies that compare the two stage and NLME approaches for analysis of experimental data. An example is given in \cite{Olofsen:2004}. Therefore one main purpose of this paper is to compare different modelling approaches, and to show that nonlinear mixed effects models and the two stage approach give different estimates of population parameters. We are also interested in investigating the parameter estimates for the SDE model, and how much that model predicts the leucine level to fluctuate during the experiment.

\subsection{Results in summary}
\subsubsection{Comparison between the two stage approach and the NLME approach}
In our study the NLME models predict a more homogeneous population. This reflects the known fact that the interindividual variability is often overestimated when the two stage approach is used \citep[see, e.g.,][]{Olofsen:2004,Sheiner&Beal:1980,Sheiner&Beal:1983,Steimer:1984}. We have also observed that, for some parameters, the two modelling approaches produced different estimates of population averages. In simulation studies with large residual errors, \cite{Sheiner&Beal:1983} have shown that the NLME approach produce less biased estimates of population averages.

\subsubsection{Using a smaller data set for each individual}
Often rich sampling of data for each individual is difficult, time consuming, or expensive. One main advantage with NLME modelling is that information from multiple individuals is used simultaneously. Therefore, population parameters can often be estimated when the number of data points is small for each individual, which may be of importance in experimental planning.

When only half of the data were used for each individual, the estimated parameters of the ODE based NLME model, where no noise inherent to the system is modelled for, were similar to when all data were used. For the SDE based NLME model the variability intrinsic to the system was estimated to be zero, indicating that data need to be sufficiently densely sampled when SDEs are used. For the two stage approach we could not estimate the parameters for the smaller data set. Thus, for the model investigated in this study, the data may be sampled less dense if the proposed ODE based NLME model is used in the analysis.

\subsubsection{Comparison between NLME models based on ODEs and SDEs}
When all parameters were allowed to vary between individuals both the NLME models based on ODEs and SDEs produced excellent fits to data and small correlations between residuals were observed. The estimated parameters were similar with the two approaches, but the interindividual variations were in general smaller for the SDE model.

The amount of free leucine in the body fluctuates over time in the SDE model. When all individual leucine levels are normalized, by dividing with the estimated average for that individual, the standard deviation of the leucine level is estimated to be $0.089$ at each time point. This means that, for a typical individual, the amount of free leucine is predicted to vary with a standard deviation of $8.9\%$ of the estimated average value, which indicate daily variations of leucine levels.

\subsubsection{Differences between populations}
For the two groups of individuals different estimations of the population medians of the parameters have been observed. Especially the level of free leucine was significantly higher for the diabetic subjects than the control individuals. Also the rate parameters differed between the two groups and the fraction of free leucine that was degraded or used for protein synthesis per time unit is smaller for the diabetic individuals. Moreover proteolysis is predicted to be larger for the diabetic individuals.

\section{Theory and methods} \label{Sec:TheoryAndMethods}
In this section we use a simple one-state compartmental model to describe the theory of compartmental models, tracer/tracee studies, and nonlinear mixed effects models. We also describe how stochastic differential equations can be used to account for a fluctuating tracee in the system. Moreover we discuss how to define an objective function to evaluate how well a model output resembles data.

\subsection{Compartmental models - Tracer studies}
A compartment is defined as a well mixed and kinetically homogeneous amount of material \citep{Cobelli:2000}. A compartmental model describes the relation between the amount of matter in a compartment and the fluxes of material in and out from the system, but also fluxes between compartments (in case of multi-compartment models).

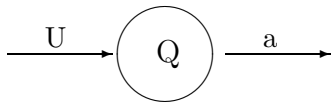
\begin{figure}[b!]
\setlength{\unitlength}{1mm}
\begin{picture}(140,14)
 \put(50,7){\vector(1,0){14}}
 \put(55,8){\text{{\small U}}}
 \put(71,7){\circle{13}}
 \put(69.8,6){Q}
 \put(79,7){\vector(1,0){14}}
 \put(84,8){\text{{\small a}}}
\end{picture}
\caption{{\footnotesize A graphical representation of a one-compartment model.}}
\label{fig:oneCompEx}
\end{figure}

Consider the one-compartment model described in Figure~\ref{fig:oneCompEx}. The influx of material into the compartment per time unit is denoted $U$, the fraction of material that leaves the compartment per time unit is denoted $a$, and the amount of material in the compartment is denoted $Q$. A mathematical representation of the model can be expressed with an ordinary differential equation,
\begin{equation}
\mathrm{d}Q/\mathrm{d}t=-aQ+U,
\end{equation}
where, in the general, representation $a$ and $U$ may be state and/or time dependent.

When the system is in steady state, i.e., when $Q$ is a constant and $0=-aQ+U$, measurements of $Q$ will give little information about the dynamical properties of the system (unless $a$ or $U$ are known, it is then enough with only one measurement point of $Q$ to describe the complete system). A common way to gain more information about the system is by tracer experiments, in which a known amount of tracer material is added into the system. Thus, a new input signal, $u=u(t)$, is introduced for the tracer, $q$, and measurements of the tracer-to-tracee ratio, $q/Q$, at successive time points may be used to get estimates of $a,\ U$, and $Q$. The tracer should have the same kinetic properties as the tracee but not affect the overall system behaviour~\citep{Anderson:1983,Cobelli:2000}.

Under the assumption that $Q$ is in steady state, the model takes the form of an ODE \mbox{$\mathrm{d}q/\mathrm{d}t=-aq+u(t)$} for the tracer, and the output (response) function, \mbox{$Y(t)=q(t)/Q$}. However, measurements are usually error-prone and it is common to add a term $v=v(t)$ to the output function, accounting for measurement errors but also for model uncertainty and noise effects intrinsic to the system. Measurements are performed at discrete time points $t_k,\ k=1,\dots,d$, and the $v(t_k)$:s are often assumed to be Gaussian (with expectation zero) and independent, i.e., $v(t_k)$ and $v(t_l)$ are independent for~$k\neq l$. When the output differs significantly between different time points an alternative is to assume a proportional error model, e.g., a lognormal distribution \mbox{$Y(t_k)=\frac{q(t_k)}{Q}\exp(v(t_k))$}. Observe that, by taking logarithms on both sides, this may be transformed to an additive noise model.

\subsection{Modelling uncertainty and intrinsic system noise} \label{Sec:InherentSystemNoise}
Stochastic differential equations may be used to account for noise effects that are inherent to the system and not due to measurement errors. Such effects may be caused by true random variations in the biological process. Moreover, in many cases the systems to be modelled are not fully understood, or too complex to be modelled exactly by systems of ODEs. The models may then be improved by adding a system noise term (turning the ODEs into SDEs) representing the lumped effect of all non-explicitly mechanistically modelled effects in the system.

In a tracer/tracee model the tracee is often assumed to be constant. However, this is not exactly true in real biological systems through the time course of an experiment, but it may vary around a mean value $Q_0$. This may, e.g., be modelled by the mean reverting Ornstein-Uhlenbeck process
\begin{equation}
\mathrm{d}Q_t=\alpha(Q_0-Q_t)\mathrm{d}t+\sigma Q_0\mathrm{d}W_t, \label{eq:TraceeOneCompModelOneIndiv}
\end{equation}
where $W_t$ is the standard Wiener process, $\sigma Q_0$ determines the magnitude of the system noise (the reason to include $Q_0$ here is explained in the population model below), and $\alpha$ is the rate at which $Q_t$ reverts towards the mean $Q_0$ (given in fraction of material per time unit). In the start of the experiment we assume that $Q$ is Gaussian with mean $Q_0$ and variance $(\sigma Q_0)^2/(2\alpha)$, which is the asymptotic mean and variance of $Q_t$ regardless of the initial value. Then the process ${Q_t}$ is stationary and Gaussian with mean \mbox{$E(Q_t)=Q_0$} and covariance \mbox{$Cov(Q_s,Q_t)=\frac{(\sigma Q_0)^2}{2\alpha}e^{-\alpha(t-s)}$} for $s \le t$ \citep{Karatzas:1991}. Thus, at each time point~$t$,
\begin{equation}
Q_t \sim N\left(Q_0,\frac{(\sigma Q_0)^2}{2\alpha}\right). \label{distr:Q_t}
\end{equation}

If there is reason to suspect that the error terms, $v(t_k)$, are not Gaussian and independent these terms may be split into two parts, $v(t_k)=v_1(t_k)+v_2(t_k)$, where $v_2$ is assumed to be Gaussian and independent, representing the measurement errors, but $v_1$ accounts for noise intrinsic to the system and may be modelled by a stochastic differential equation, e.g., a zero-mean Ornstein-Uhlenbeck process \citep{Moller:2010}.

Other approaches to account for system noise and uncertainty are to allow other parameters to vary randomly over time or to lump the noise and uncertainty into additive terms in the right hand side of the ODEs. Assuming these additive terms to be stochastic processes turns the ODEs into stochastic differential equations.

\subsection{Nonlinear mixed effects models with stochastic differential equations applied to tracer/tracee experiments} \label{Sec:Theory-NLME_SDE}
When time series data come from a number of individuals in a population, the model parameters are traditionally estimated separately for each individual in the studied population (independently of population knowledge). One can then compute population averages and other statistical properties from the parameters obtained for each individual. This approach to population modelling is called the ``two stage approach'' \citep{Sheiner&Beal:1980} or the ``standard two stage approach'' \citep{Sheiner&Beal:1983,Steimer:1984}.

Another approach to model population behaviour is the statistical framework of nonlinear mixed effects modelling. NLME models are hierarchical in structure with a clear separation of the intraindividual and interindividual variations. The interindividual differences are characterized by individual-specific values of the parameters, for which the distribution of the population is assumed to be known. The intraindividual variation is defined as the individual measurement errors and system noise.

In Equations~\mbox{\eqref{eq:TraceeOneCompModel}-\eqref{eq:Q0iOneComp}} we use the simple one-compartment model to illustrate how NLME models with SDEs may be constructed. It is a model where the tracee fluctuates around a mean steady value. The interindividual variations are seen in~\eqref{eq:aiOneComp} and~\eqref{eq:Q0iOneComp}, whereas the intraindividual variations are accounted for in Equations~\eqref{eq:TraceeOneCompModel} and~\eqref{eq:outputOneCompModel}. The population model takes the form
\begin{align}
\mathrm{d}Q_t^i & = \alpha(Q_0^i-Q_t^i)\mathrm{d}t+\sigma Q_0^i\mathrm{d}W_t, & Q^i(0) & \sim N\left(Q_0^i,\frac{(\sigma Q_0^i)^2}{2\alpha}\right), \label{eq:TraceeOneCompModel}\\
\mathrm{d}q_t^i & = (-a^iq_t^i+u^i(t))\mathrm{d}t, & q^i(0) & = 0, \label{eq:TracerOneCompModel} \\
Y_k^i           & = \log(\frac{q_k^i}{Q_k^i}) + v_k^i, & & \label{eq:outputOneCompModel} \\
a^i             & = a\exp(\eta_a^i), & & \label{eq:aiOneComp} \\
Q_0^i           & = Q_0\exp(\eta_Q^i), & & \label{eq:Q0iOneComp}
\end{align}
where $i=1,\dots,N$ denotes the individual, subscript $k=1,\dots,d^i$ denotes the variable value at sample time point $t_k$ for the $i$:th individual, and the measurement errors, $v_k^i$, are assumed to be independent and Gaussian with expectation zero and variance $S_k$. The measurements have been log-transformed which results in an additive noise structure in~\eqref{eq:outputOneCompModel}.

The initial condition of the tracee state variable in Equation~\eqref{eq:TraceeOneCompModel} is the asymptotic distribution of $Q_t^i$. If the tracer material is introduced into the system as a bolus input at time zero, an impulse function can be used for the tracer input function $u^i(t)$. A mathematically equivalent model would be to use $u^i=0$ and \mbox{$q^i(0)=q^i_0$}, where $q^i_0$ is the amount of introduced tracer material. The reason for using $\sigma Q_0^i$ as a diffusion constant in~\eqref{eq:TraceeOneCompModel} is that the level of $Q_t^i$ differs between individuals.

In~\eqref{eq:aiOneComp} and~\eqref{eq:Q0iOneComp} $a^i$ and $Q_0^i$ depend on the fixed effects parameters $a$ and $Q_0$, thought to represent a typical individual in the population, and the individual-specific random parameters \mbox{$\eta^i=[\eta_a^i,\eta_Q^i]^T$}. In this paper the latter are assumed to be Gaussian over the population with expectation zero and covariance matrix \mbox{$\Omega=\text{diag}(\omega_a^2,\omega_Q^2)$}. Lognormal distributions are used for $a^i$ and $Q_0^i$ mainly since it prevents parameters from being negative, even if the interindividual variabilities are large. The parameters $a$ and $Q_0$ do not represent means, but instead the medians of the lognormal distribution \citep{Blackwood:1992}.

Note that the model defined by \mbox{\eqref{eq:TraceeOneCompModel}-\eqref{eq:Q0iOneComp}} is similar to the pharmacokinetics NLME model with SDEs in \cite{Overgaard:2005}. There the plasma volume corresponds to the tracee variable, and is assumed to be constant. Noise is added to the differential equation that describes the kinetics of the drug, corresponding to the tracer Equation~\eqref{eq:TracerOneCompModel} in the above model.

\subsection{Approximation of the population likelihood function} \label{Sec:APLapprox}
For parameter estimation, we apply the maximum likelihood framework where the goal is to find the parameter set that maximizes the likelihood function, defined as the probability of finding the measured output for a given set of parameters. In the presentation below we allow for multiple measurement variables, in contrast to in~\eqref{eq:outputOneCompModel} where we only have one measured variable.

To construct a likelihood function we assume that the densities in the output function are Gaussian at all time points, $t_k$, and dependent only on the information available at time $t_{k-1}$. For \mbox{$k=2,\dots,d^i$}, let \mbox{$\mathcal{Y}^i_{k-1}=[y_1^i,\dots,y_{k-1}^i]^T$} be the data up to time $t_{k-1}$ for individual $i$. Let
\begin{equation} \label{eq:predictions}
Y^i_{k|k-1}=E(Y^i_k|\mathcal{Y}^i_{k-1})
\end{equation}
be the predicted mean value at time $t_k$, conditioned on the information available at time $t_{k-1}$, and
\begin{equation} \label{eq:residualCovariance}
R_k^i=R_{k|k-1}^i=Cov(Y_k^i|\mathcal{Y}^i_{k-1}),
\end{equation}
the corresponding conditional covariance matrix. Then the residual vector \mbox{$\epsilon_k^i=y_k^i-Y_{k|k-1}^i$} is Gaussian with expectation zero and covariance $R_k^i$. The starting point, $(Y_{1|0}^i,R_1^i)$, for the recursion algorithm needed to compute~\mbox{\eqref{eq:predictions}-\eqref{eq:residualCovariance}} is determined by the initial distributions of the state variables.

Kalman filtering techniques may be used to compute $\epsilon_k^i$ and $R_k^i$. The Kalman filter (KF) can be applied if the model is linear. It is optimal in the sense that it minimizes the variance of the prediction error if it is assumed that the densities of the model outputs and the initial conditions are Gaussian, and that the covariance matrix for the Gaussian noise in the state equations and the output noise covariance are state independent~\citep{Jazwinski:1970}. The basic idea is that at every sample the conditional distributions of the state variables are updated by taking the new data point into account. This typically results in a net movement from the predicted mean value of the state in the direction of the experimental observations. For nonlinear models, such as the one in Equation~\eqref{eq:outputOneCompModel}, the extended Kalman filter may be used \citep{Kristensen:2004,Overgaard:2005,Jazwinski:1970}. It is based on a linearization (in each time step,$t_k$) of the nonlinear parts of the model, which results in approximate equations to which the KF can be applied.

The population likelihood function is a marginal likelihood function, where all the individual parameters, $\eta^i$, are integrated out. The Gaussian densities of the residuals and the individual-specific parameters, $\eta^i$, gives a likelihood function of the form (see, e.g., \citealt{Overgaard:2005})
\begin{equation}\label{Eq:popLikelihood}
L(\theta;\mathcal{Y}) = \prod^N_{i=1}\int_{\mathbb{R}^r} \exp(l^i)\mathrm{d}\eta^i,
\end{equation}
where $\theta$ is the vector of population parameters (i.e., \mbox{$\theta=[a,Q_0,\omega_a^2,\omega_Q^2,\sigma,S]^T$} in model~\mbox{\eqref{eq:TraceeOneCompModel}-\eqref{eq:Q0iOneComp}}), $\mathcal{Y}=\{\mathcal{Y}^1,\dots,\mathcal{Y}^N \}$ is the complete set of measurement data. The individual log-likelihood function $l^i=l^i(\eta^i)=l^i(\eta^i;\mathcal{Y}^i,\theta)$ is given by
\begin{equation}\label{Eq:indLikelihood}
l^i = -\frac{1}{2}\Big(\sum^{d^i}_{k=1}\Big[{\epsilon^i_k}^T{R^i_k}^{-1}\epsilon^i_k+\log|R^i_k|+m\log(2\pi)\Big] + {\eta^i}^T\Omega^{-1}\eta^i+\log|\Omega|+r\log(2\pi){}\Big),
\end{equation}
where $m$ is the number of outputs and $r$ is the number of interindividual random effects, i.e., the dimension of $\eta^i$.

In general it is not possible to find an exact expression of the population likelihood function for a model with SDEs. The integrand must therefore be approximated and we have applied the Laplace approximation method together with the first-order conditional estimation (FOCE) method (see, e.g., \citealt{Overgaard:2005,Wang:2007}). The Laplace approximation method is based on a second order truncation of the Taylor expansion of $l^i$ around a stationary point $\hat{\eta}^i$. In the FOCE method $\hat{\eta}^i$ is estimated as
\begin{equation}
\hat{\eta}^i=\argmin_{\eta^i}\left\{ -l^i(\eta^i;\mathcal{Y}|\theta)\right\},\label{eq:etaOptFunc}
\end{equation}
in contrast to the first order (FO) method that uses $\hat{\eta}^i=0$ (see, e.g., \citealt{Overgaard:2005}, for the details). With the Laplace approximation method the population likelihood function in \eqref{Eq:popLikelihood} is approximated as
\begin{equation}\label{eq:popLikelihoodApprox}
L(\theta;\mathcal{Y}) \approx
\prod^N_{i=1}\exp(l^i(\hat{\eta}^i)) \Big| \frac{H_\eta(l^i(\hat{\eta}^i))}{(2\pi)^r} \Big|^{-1/2},
\end{equation}
where, by disregarding second order derivatives of the residuals, $\epsilon_k^i$, and covariance matrices, $R_k^i$, the Hessian $H_\eta(l^i(\hat{\eta}^i))$ is approximated as
\begin{align}
\left(H_\eta(l^i(\hat{\eta}^i))\right)_{a,b} = \frac{\partial^2(l^i(\hat{\eta}^i))}{\partial\eta_a\partial\eta_b} & \approx -\sum_{k=1}^{d_i}\frac{\partial{\epsilon^i_k}^T}{\partial\eta_a}{R^i_k}^{-1}\frac{\partial\epsilon^i_k}{\partial\eta_b} \nonumber \\
&+\sum_{k=1}^{d_i}\left({\epsilon^i_k}^T{R^i_k}^{-1}\frac{\partial R^i_k}{\partial\eta_a}{R^i_k}^{-1}\frac{\partial\epsilon^i_k}{\partial\eta_b}+\frac{\partial{\epsilon^i_k}^T}{\partial\eta_a}{R^i_k}^{-1}\frac{\partial R^i_k}{\partial\eta_b} {R^i_k}^{-1}\epsilon^i_k\right) \nonumber \\
&-\sum_{k=1}^{d_i}{\epsilon^i_k}^T{R^i_k}^{-1}\frac{\partial R^i_k}{\partial\eta_a}{R^i_k}^{-1}\frac{\partial R^i_k}{\partial\eta_b}{R^i_k}^{-1}\epsilon^i_k \label{eq:indHessianApprox} \\
&-\frac{1}{2}\sum_{k=1}^{d_i}\mathrm{Tr}\left(-{R^i_k}^{-1}\frac{\partial R^i_k}{\partial\eta_a}{R^i_k}^{-1}\frac{\partial R^i_k}{\partial\eta_b}\right) \nonumber \\
&-\left(\Omega^{-1}\right)_{a,b}, \nonumber
\end{align}
in position $(a,b),\ a,b=1,\dots,r$. Proving~\eqref{eq:indHessianApprox} is elementary by using differentiation rules for matrices found in \cite{MatrixCookbook:2008}. Differentiation with respect to a matrix (or vector) in~\eqref{eq:indHessianApprox} is performed on all elements in the matrix. If there is no interaction between the interindividual effects, $\eta^i$, and the intraindividual variation effects, the derivatives of $R^i_k$ are zero and \mbox{$\left(H_\eta(l^i(\hat{\eta}^i))\right)_{a,b} \approx -\sum_{k=1}^{d_i}\frac{\partial{\epsilon^i_k}^T}{\partial\eta_a} {R^i_k}^{-1} \frac{\partial\epsilon^i_k}{\partial\eta_b}-\left(\Omega^{-1}\right)_{a,b}$}.

Every separate evaluation of the approximate population likelihood function requires the estimation of the individual parameters $\eta^i$ when the FOCE method is applied. This may result in computationally heavy calculations, especially when the number, $r$, of interindividual effects is large.

\subsection{Implementation} \label{Sec:Implementation}

All the implementations have been performed in MATLAB$^{\textregistered}$ (2008b, The MathWorks, Natick, Massachusetts, USA).

\subsubsection{Optimization methods}
For NLME models the population parameters are estimated as
\begin{equation}\label{eq:popOptFunc}
\hat{\theta}=\argmin_{\theta}\left\{-\log(L(\theta;\mathcal{Y}))\right\}.
\end{equation}

To optimize~\eqref{eq:etaOptFunc} we have used a quasi-Newton method based on the so called BFGS (Broyden, Fletcher, Goldfarb, and Shanno) updating formula \citep{Nocedal:1999}. To find the optimal population parameters in~\eqref{eq:popOptFunc} we start with the derivative free Nelder-Mead simplex-reflection method \citep{Nocedal:1999} for a finite number of iterations to get closer to the true optimum. Then the optimization has been fine-tuned by the BFGS-method. The gradients used in the BFGS method have been approximated by a finite central-difference approximation.

The Nelder-Mead simplex-reflection method and the BFGS method are both local optimization methods, i.e., only local minima of the objective functions are searched for. However, for the models investigated in this paper, different initial guesses of the parameters always resulted in convergence to the same optimal parameter set, indicating that the objective functions do not have multiple local minima.

The steps that we have used to estimate the parameters in the NLME settings are:
\begin{flushleft}
\begin{itemize}
\item Choose an initial guess, $\theta_0$, of the population parameters and $\eta_0^i=0$ for the individual parameters ($i=1\dots,N$).
\item Calculate the population likelihood~\eqref{eq:popLikelihoodApprox} by optimizing $\eta^i$ in~\eqref{eq:etaOptFunc} for $i=1,\dots,N$. Let $\hat{\eta}_0^i$ denote the optimized $\eta^i$:s.
\item Set $j=1$ and repeat the following until a stopping criterion is reached:
\begin{itemize}
\item[-] Find a new $\theta_j$ according to the optimization algorithm by using $\hat{\eta}_{j-1}^i$ as initial values in every new calculation of~\eqref{eq:popLikelihoodApprox}.
\item[-] Denote $\hat{\eta}_j^i$ the optimized $\eta^i$ for $\theta=\theta_j$.
\item[-] Set $j=j+1$.
\end{itemize}
\end{itemize}
\end{flushleft}

\subsubsection{Inference on estimated parameters}
For inference on the estimated parameters of an NLME model an approximation of the covariance matrix may be obtained from
\begin{equation} \label{eq:covarianceMatrix}
Cov(\hat{\theta})\approx(H_\theta(-\log(L(\hat{\theta};\mathcal{Y}))))^{-1}
\end{equation}
\citep{Kristensen:2004}, where $H_\theta$ denotes the Hessian with respect to $\theta$. The maximum likelihood estimate is Gaussian and the covariance matrix is the inverse of the Fisher information matrix, which is the expected value of the Hessian of the negative log-likelihood function when data are seen as random variables \citep{Pawitan:2001}. For a given realization, $\mathcal{Y}$, of the data, the \emph{observed} Fisher information matrix, \mbox{$H_\theta(-\log(L(\hat{\theta};\mathcal{Y})))$,} may be used instead \citep{Pawitan:2001}. However, in this study we only approximate the population likelihood function in~\eqref{eq:popLikelihoodApprox} and the Hessian is computed numerically by means of a central difference procedure around the estimated parameters.

When different mathematical models are used it is interesting to test if they give significantly different estimated values of the same physical quantities. With the NLME approach we approximate the confidence intervals around each estimated parameter by using~\eqref{eq:covarianceMatrix} and assuming that $\hat{\theta}$ is Gaussian. Confidence intervals around the estimated population parameters can also be calculated by taking every individual into account, as discussed in Section~\ref{Sec:leucine model - parEst}.

If the confidence intervals around a parameter are disjoint when two different models are used, we say that the models predict the parameter differentially with statistical significance at a certain confidence level. In a study where sample means from repeated measurements from normally distributed data were compared, \cite{Payton:2000} suggest to use $85\%$ confidence intervals to test the hypothesis that the means are equal at significance level $0.05$. We follow that suggestion when we compare the parameters of our models, but it should be mentioned that the confidence intervals in our study are not calculated from repeated realizations from a normal distribution.

\subsubsection{The extended Kalman filter}
The extended Kalman filter has been implemented as described in \cite{Kristensen:2004}. The EKF needs estimates of the predicted initial state and the initial state covariance matrix. As discussed in Section~\ref{Sec:InherentSystemNoise}, we use \mbox{$Var(Q(0))=(\sigma Q_0^i)^2/(2\alpha)$} as an estimate of the variance of the tracee state variable modelled as in Equation~\eqref{eq:TraceeOneCompModel}. When the underlying model is deterministic in the state variables (ODE based), the state covariance is zero.

The state equations for the models examined in this study are linear differential equations and the only non-linear parts of the models are present in the model output equations. Therefore the linearization approximation in the EKF is only applied to the model output equation.

\section{Population Models of Leucine Kinetics}
In this article NLME models based on ordinary and stochastic differential equations are constructed for a compartmental model that describes the kinetics of leucine in blood plasma. The ODE based model has been used in several publications as a submodel of larger compartmental models that describe the kinetics of low density lipoproteins in plasma, see, e.g., \cite{Adiels:2005_2,Demant:1996}, or the review in \cite{Barrett:2006}. As far as we know the model parameters are estimated on single individuals and population features have been inferred with the two stage approach in all studies up to date.

When the leucine model has been applied to fit parameters to data from single individuals the estimated values of the parameters vary much over the population. An explanation for this may be that measurement errors in data are not taken care of \citep{Olofsen:2004}. One benefit with the NLME approach is that it handles intraindividual variations explicitly. In our implementation large interindividual variations are penalized since the individual log-likelihood in Equation~\eqref{Eq:indLikelihood} becomes smaller for larger absolute values of the elements in $\eta^i$ (through the term $-{\eta^i}^T\Omega^{-1}\eta^i$).

A stochastic differential equation is used to account for tracee deviations from steady state as in Equation~\eqref{eq:TraceeOneCompModel}. It should be pointed out that a population approach such as NLME modelling is needed to estimate parameters in SDEs when the data are sparsely sampled for each individual, since rich sampling is needed to separate system noise from measurement noise in single subject estimation algorithms \citep{Overgaard:2005}. A population approach such as an NLME model allows for more data to be used simultaneously.

\begin{figure}[b!]
\begin{center}
\includegraphics[width=10cm,height=6cm]{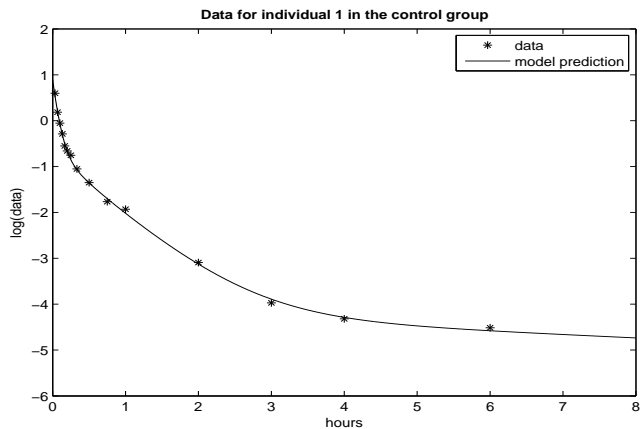}
\end{center}
\caption{{\footnotesize Logarithm of the data for an individual in the control group. The output predicted by the ODE based NLME model described in Section~\ref{Sec:leucine model - pop} is also included in the figure.}}
\label{fig:dataInd1}
\end{figure}

The data used are taken from experiments where stable isotope-labelled leucine ($^2H_3$-leucine) was introduced in the system as a bolus input. The experimental setup is described in \cite{Adiels:2005_1,Adiels:2005_2,Adiels:2006}. The data set is divided into one group of control individuals and one group of individuals suffering from diabetes mellitus type~$2$. The data from the~$19$ individuals in the control group are also used in \cite{Adiels:2005_1,Adiels:2005_2,Adiels:2006}. The data from the~$15$ diabetic individuals are also used in \cite{Adiels:2005_1,Adiels:2006}. Each individual was fasting $12$ hours prior to the tracer injection and also during the sampling period. The samples were collected~$16$ times at $2,\ 4,\ 6,\ 8,\ 10,\ 12,\ 15,\ 20,\ 30$, and $45$ minutes and $1,\ 2,\ 3,\ 4,\ 6$, and $8$ hours after the tracer injection. However there are missing data for some individuals. The data for a single control individual is visualized in Figure~\ref{fig:dataInd1}.

We will now describe the ODE based individual model (additive Gaussian measurement noise) used in a two stage approach to indirectly compute population parameter properties (Section~\ref{Sec:leucine model}). Then we present the ODE based NLME model (additive Gaussian measurement noise, lognormal distributed population parameters) used directly to estimate parameter values and their distributions taking a maximum likelihood approach (Section~\ref{Sec:leucine model - pop}). Thereafter the SDE based NLME model (additive system noise) is discussed (Section~\ref{Sec:leucine model - popSDE}). The estimated population parameters are presented after each model has been presented. Finally we compare the results for the NLME models based on ODEs and SDEs (Section~\ref{Sec:ODEvsSDE}).

\subsection{The ODE based individual model used in a two stage approach} \label{Sec:leucine model}
Leucine is an essential amino acid, which means that it is not synthesized in the body and that humans must acquire it from the diet. It is an important element in apolipoprotein~B (apoB), and since low density lipoproteins contain exactly one apolipoprotein B-100 molecule, stable isotope-labelled leucine can be used as tracer in kinetic studies of low density lipoproteins. The model used in this article describes the kinetics of leucine in plasma before it enters the apoB synthesis machinery in the liver. The model was developed by \cite{Demant:1996}, and is based on the work in \cite{Cobelli:1991}. The data consist only of measurements of free leucine in plasma and is expressed as a tracer-to-tracee mass ratio. Figure~\ref{fig:leuModel} shows a schematic view of the model.

Compartment~$1$ corresponds to the amount of free leucine in the plasma. It has a continuous inflow of leucine from other parts of the body and also an exit (corresponding to catabolism of leucine and uptake of leucine by body proteins with a very slow turnover rate), described by the fractional transfer coefficient parameter $k_{01}$. The amount of inflow of unlabelled free leucine per time unit is given by the parameter $U_{1}$. Since the tracer was introduced as a bolus input all the labelled leucine is assumed to be in compartment~$1$ at time zero. No additional labelled leucine is assumed to be introduced into the system. Compartments~$3$ and~$4$ correspond to a body protein pool that account for the uptake and subsequent slow release of leucine back to the plasma. Compartment~$2$ is a hepatic intracellular compartment. From this compartment the liver is fed with leucine that is used when apolipoproteins are synthesized. The parameter that describes the fraction of material in compartment~$2$ that reach the apoB synthesis machinery in the liver per time unit is here denoted $k_{L2}$. When time passes the system is washed out from labelled material and the amount of tracer in the system converges to zero. Throughout this paper the time unit used is hours, thus all fractional transfer coefficients are presented in the unit $h^{-1}$. The amount of material in compartments is presented in mg.

The ordinary differential equations in the model of the tracee system are \citep{Adiels:2005_2,Demant:1996}
\begin{align}
\mathrm{d}Q_1/\mathrm{d}t & = -(k_{01}+k_{21}+k_{31})Q_1 + k_{12}Q_2 + k_{13}Q_3 + U_1 \label{eq:Qprim1} \\
\mathrm{d}Q_2/\mathrm{d}t & = -(k_{12}+k_{L2})Q_2 + k_{21}Q_1 \label{eq:Qprim2} \\
\mathrm{d}Q_3/\mathrm{d}t & = -(k_{13}+k_{43})Q_3 + k_{31}Q_1 + k_{34}Q_4 \label{eq:Qprim3} \\
\mathrm{d}Q_4/\mathrm{d}t & = -k_{34}Q_4 + k_{43}Q_3, \label{eq:Qprim4}
\end{align}
where $Q_j=Q_j(t)$ is the amount of tracee material in compartment $j$ at time $t$. We assume here that the fractional transfer coefficients are not state dependent. Hence the system is linear, and the systems for the tracer and the tracee are identical except for $U_1$ which is controlled by the experimentalist for the tracer system (it is assumed that no labelled material is produced in the body).

Let \mbox{$Q=[Q_1,Q_2,Q_3,Q_4]^T$} and \mbox{$q=[q_1,q_2,q_3,q_4]^T$} be state vectors for the tracee- and tracer system respectively, and \mbox{$U=[U_1,0,0,0]^T$} be a vector of input variables in the tracee system. Then the complete model can be written compactly in matrix notation as
\begin{align}
\mathrm{d}Q/\mathrm{d}t & = KQ + U \label{eq:Tracee} \\
\mathrm{d}q/\mathrm{d}t & = Kq \label{eq:Tracer} \\
Y_k & = q_1(t_k)/Q_1(t_k), \label{eq:Output}
\end{align}
\noindent
where \eqref{eq:Tracee} corresponds to the tracee system, \eqref{eq:Tracer} to the tracer system, and \eqref{eq:Output}
corresponds to model output taken at discrete time points $t_k,\ k=1,\dots,d$, where $d$ is the number of samples. Observe that the experiments are based on a bolus injection in compartment $1$ at time zero. It is assumed that the tracer material is instantly well stirred throughout the plasma and that there is no tracer material in the system before the experiments start, therefore $q(0)=[q_{10},0,0,0]^T$, where $q_{10}$ is the known amount of labelled leucine injected ($7$ mg/[kg body weight] in our case). Another assumption is that the exchange of hydrogen atoms between leucine molecules and other molecules in the body is negligible, i.e., it is assumed that labelled leucine molecules remain labelled throughout the experiment. The matrix $K$ is the compartmental matrix of the system, i.e., the coefficient matrix of the linear system~\mbox{\eqref{eq:Qprim1}-\eqref{eq:Qprim4}}.

The tracee system is approximated by a steady state \citep{Adiels:2005_2}, i.e.,~\mbox{$\mathrm{d}Q/\mathrm{d}t=0$} and the fractional transfer coefficients, $k_{ij}$, are constant. In matrix notation we get \mbox{$0=KQ+U \Rightarrow U=-KQ$} or \mbox{$Q=-K^{-1}U$} for the tracee equations. Compartmental matrices are always invertible if there are at least one exit compartment (a compartment with a flow out of the system) and the system contains no traps (subsystems with no flows to the outside of the subsystem) \citep{Cobelli:2000}.

When the leucine system is considered separately (without the apoB part) the fractional rate, $k_{L2}$ (per hour), of material leaving compartment~$2$ for lipoprotein synthesis is unidentifiable and in this article it is set to $0.01h^{-1}$. This value has been chosen because it is the value that we have found for most subjects in an implementation of the full apoB-model. Further parameter constraints have been used to reduce the number of unknowns. This is actually necessary to get an a priori identifiable model since, without the constraints, three different parameter sets describe the same model structure \citep{Cobelli:1980}. We have followed \cite{Packard:2000} and used $k_{21}=k_{12}$ (the rate of the flow of leucine through the cell walls of liver cells are the same in both directions) and $k_{34}=0.1k_{43}$. It is not necessary to estimate both $U_1$ and $Q_1$ separately since \mbox{$U=-KQ \Rightarrow U_1=Q_1(k_{01}(k_{12}+k_{L2})+k_{21}k_{L2})/(k_{12}+k_{L2})$} mg/hour. Thus, the parameters that are estimated for a single individual are $k_{01},\ k_{12},\ k_{13},\ k_{31},\ k_{43}$, and $Q_1$.

\subsubsection{Estimated parameters} \label{Sec:leucine model - parEst}
Assuming a steady state for the tracee, the parameters were estimated for each individual separately. We used the weighted least squares method, taking the measured values as weights at each time point.

\begin{table}[b!]
\begin{center}
\begin{scriptsize}
\begin{tabular}{|c|c|c|c|c|c|c|}
\hline
 &$\mathbf{k_{01}}$&$\mathbf{k_{12}}$&$\mathbf{k_{13}}$&$\mathbf{k_{31}}$&$\mathbf{k_{43}}$&$\mathbf{Q_1}$ \\\hline
\textbf{All individuals}&2.15 (41$\%$)&1.77 (116$\%$)&1.78 (115$\%$)&3.48 (70$\%$)&1.00 (63$\%$)&373  (65$\%$) \\\hline
\textbf{Controls}&2.32 (32$\%$)&1.70 (137$\%$)&1.61 (132$\%$)&4.18 (37$\%$)&0.94 (53$\%$)&289 (43$\%$) \\\hline
\textbf{Diabetics}&1.96 (51$\%$)&1.86 (98$\%$)&2.04 (95$\%$)&2.76 (98$\%$)&1.09 (80$\%$)&516  (73$\%$) \\\hline
\end{tabular}
\end{scriptsize}
\caption{{\footnotesize Estimated geometric mean (and coefficient of variation) of the parameters when data from the $34$ individuals are used. The population parameters are trimmed as explained in the text.}}
\label{table:results_STS}
\end{center}
\end{table}

The geometric mean and the coefficients of variation (CoV) of the full population are presented in Table~\ref{table:results_STS} for the six parameters. The reason for reporting the geometric mean instead of the arithmetic mean is that the parameters are approximately lognormally distributed in the population. Therefore, the variations of the parameters are multiplicative. The geometric mean is a maximum likelihood estimator of the median of a lognormal distribution \citep{Blackwood:1992}. The coefficient of variation of the multiplicative model is a normalized measure of the spread of the parameters. It relates the standard deviation to the median of the parameters and is calculated by $\text{CoV}=100\times\sqrt{\exp(s^2)-1}\ \%$, where $s^2$ is the unbiased estimate of the variance of the logarithm of the estimated parameters \citep{Blackwood:1992}.

The parameters $k_{12}$ and $k_{13}$ are the ones that display the largest variation over the population. This is, at least partially, because the objective function is less sensitive to variations in these parameters. Therefore they are more influenced by noise in the observations than the other parameters.  The estimated population parameters are trimmed in the sense that a few estimates are manually removed from the calculations (e.g., $k_{01}$ is estimated to be $0.001$ - the predefined lower bound - for two individuals). All parameters except $Q_1$ had a few individual estimates that were manually removed.

To compare the results from the two stage approach with the results from the NLME approach in the forthcoming sections we calculated $85\%$ confidence intervals for the medians of the parameters for respective group (see Figure~\ref{fig:CIs}, where the corresponding intervals for the NLME models are also included). We especially observe that, for the two stage estimates, $Q_1$ differs between control group $1$ and the diabetic group. The confidence intervals have been calculated as explained below.

If a parameter $p$ is normally distributed over the population with unknown mean and variance, a $100(1-\hat{\alpha})\%$ confidence interval of the population mean of $p$ can be estimated as \mbox{$\bar{p} \pm t_{\hat{\alpha}/2,N-1}\frac{s}{\sqrt{N}}$}, where $\bar{p}$ is the sample mean of the $N$ observations of $p$, $s$ the (unbiased) sample standard deviation of the observations, and $t_{\hat{\alpha}/2,N-1}$ is the $100(1-\hat{\alpha}/2)$ percentile of the $t$-distribution with $N-1$ degrees of freedom \citep{Larsen:2001}. However, since the parameters are assumed to be lognormally distributed over the population we calculated confidence intervals $(a,b)$ for the means of the logarithms of the estimated parameters and used $(e^a,e^b)$ to obtain confidence intervals of the medians of the parameters as suggested in \cite{Blackwood:1992}.

To compare the interindividual variations for the two stage approach and the NLME models in Sections~\ref{Sec:leucine model - pop} and~\ref{Sec:leucine model - popSDE} we have also computed confidence intervals of the population variance of $\log(\tilde{p}^i)$, where $\tilde{p}^i$ for an individual $i$ is its actual parameter estimate $p^i$ divided by the group geometric mean of $p$ for that individual. The confidence interval around a variance, $\tilde{\sigma}^2$, can be calculated since $\frac{(N-1)\tilde{s}^2}{\tilde{\sigma}^2}$ is $\chi^2$-distributed with $N-1$ degrees of freedom if the sample comes from a normal distribution \citep{Larsen:2001}. We assume that $\log(\tilde{p}^i),\ i=1,\dots,N$, is normally distributed and the estimated $85\%$ confidence interval for the interindividual variations are presented in Figure~\ref{fig:CIs}.

\subsection{The ODE based NLME model} \label{Sec:leucine model - pop}
Equations~\eqref{eq:Tracee}-\eqref{eq:Output} and the above discussion lead to the following population model,
\begin{align}
\mathrm{d}q^i_t/\mathrm{d}t & = K^iq^i_t, & q^i_0 & =[q_{1,0}^i,0,0,0]^T, \label{eq:stateNLME_ODE} \\
Y^i_k                       & = \log(\frac{q^i_{1,k}}{Q^i_1}) + v_k^i, & v_k^i & \sim N(0,S), \label{eq:OutputWithNoise}
\end{align}
where superscript $i=1,\dots,N$ denotes the $i$:th individual. When considering one individual, this is the same model as the one developed in Section~\ref{Sec:leucine model}, except that an error term, $v_k^i$, is added to the output function. The errors $\{v_k^i\}_{k=1}^{d_i}$ are assumed to be a sequence of independent normally distributed random variables with variance $S$.  In this model structure they encompass measurement errors, as well as model specification errors and system stochasticity. When considering the whole population the main differences lie in $K^i$, whose coefficients are
\begin{align}
k_{01}^i & = (G_C^ik_{01}^C+G_D^ik_{01}^D)\exp(\eta_{01}^i), & \eta_{01}^i & \sim N(0,\omega_{01}^2), \label{eq:k01i} \\
k_{12}^i & = (G_C^ik_{12}^C+G_D^ik_{12}^D)\exp(\eta_{12}^i), & \eta_{12}^i & \sim N(0,\omega_{12}^2), \label{eq:k12i} \\
k_{L2}^i & = 0.01, \label{eq:kL2} \\
k_{21}^i & = k_{12}^i , \label{eq:k21i} \\
k_{13}^i & = (G_C^ik_{13}^C+G_D^ik_{13}^D)\exp(\eta_{13}^i), & \eta_{13}^i & \sim N(0,\omega_{13}^2), \label{eq:k13i} \\
k_{31}^i & = (G_C^ik_{31}^C+G_D^ik_{31}^D)\exp(\eta_{31}^i), & \eta_{31}^i & \sim N(0,\omega_{31}^2), \label{eq:k31i} \\
k_{43}^i & = (G_C^ik_{43}^C+G_D^ik_{43}^D)\exp(\eta_{43}^i), & \eta_{43}^i & \sim N(0,\omega_{43}^2), \label{eq:k43i} \\
k_{34}^i & = 0.1k_{43}^i, \label{eq:k34i}
\end{align}
and in $Q_1^i$, which is given by
\begin{align}
Q_1^i    & = (G_C^iQ_1^C+G_D^iQ_1^D)\exp(\eta_{Q_1}^i),   & \eta_{Q_1}^i & \sim N(0,\omega_{Q_1}^2). \label{eq:Q1i}
\end{align}
The group index variable $G_C^i$ (and $G_D^i$) equals one if individual $i$ belongs to the control group (diabetic group) and zero otherwise. The fixed effects parameters $k_{01}^j$, $k_{12}^j$, $k_{13}^j$, $k_{31}^j$, $k_{43}^j$, and $Q_1^j$ are the same for all individuals in group $j$ and represent the parameters for a typical individual in respective group (where $j$ is $C$ or $D$). The parameters are assumed to have a lognormal distribution in the population to avoid negative parameter values.

To sum up, the unknown population parameters that may be estimated in the ODE based NLME framework are: the structural population fixed effects parameters $k_{01}^j$, $k_{12}^j$, $k_{13}^j$, $k_{31}^j$, $k_{43}^j$, and $Q_1^j$ (for $j=C,D$), the variances $\omega_{01}^2,\ \omega_{12}^2,\ \omega_{13}^2,\ \omega_{31}^2,\ \omega_{43}^2$, and $\omega_{Q_1}^2$ of the interindividual variation, and the variance $S$ of the measurement error term. Observe that we assume that the interindividual and the intraindividual variations are the same for the two groups in this model.

The results from the two stage approach is used as an initial guess of the structural parameters. The initial guess for $S$ was obtained by first estimating $S$ alone, with all the other parameters kept fixed to their initial values.

\subsubsection{Estimated parameters} \label{Sec:leucine model - pop - parEst}
Table~\ref{table:results_NLME_expData} shows the estimated parameters and standard errors (SEs) for the ODE model \mbox{\eqref{eq:stateNLME_ODE}-\eqref{eq:Q1i}} applied to the experimental data (the estimated parameters from the SDE model explained in the next section are also included in the table). Observe that the uncertainties of the $\omega$-parameters are large compared to the estimated values.

\begin{table}[b!]
\begin{center}
\begin{footnotesize}
\begin{tabular}{|c|c|c|c|c|c|c|}
\hline
\multicolumn{1}{|c}{} & \multicolumn{2}{|c|}{\textbf{ODE - all data}} & \multicolumn{2}{|c|}{\textbf{ODE - less data}} & \multicolumn{2}{|c|}{\textbf{SDE - all data}}\\\hline
& \textbf{Controls}  & \textbf{Diabetics} & \textbf{Controls}  & \textbf{Diabetics} & \textbf{Controls}  & \textbf{Diabetics} \\\hline
$\mathbf{k_{01}}$ & 2.58 (0.13) & 2.08 (0.12) & 2.60 (0.13) & 2.05 (0.13) & 2.64 (0.14) & 2.06 (0.12) \\\hline
$\mathbf{k_{12}}$ & 1.75 (0.16) & 1.69 (0.20) & 1.51 (0.27) & 1.78 (0.22) & 1.61 (0.16) & 1.56 (0.20) \\\hline
$\mathbf{k_{13}}$ & 3.74 (0.54) & 5.47 (0.95) & 3.95 (0.57) & 4.93 (1.11) & 4.00 (0.50) & 4.82 (0.76) \\\hline
$\mathbf{k_{31}}$ & 3.96 (0.39) & 3.40 (0.44) & 4.24 (0.44) & 2.88 (0.51) & 4.34 (0.48) & 3.32 (0.47) \\\hline
$\mathbf{k_{43}}$ & 1.22 (0.17) & 1.38 (0.26) & 1.06 (0.23) & 1.54 (0.34) & 1.16 (0.14) & 1.27 (0.22) \\\hline
$\mathbf{Q_1}$    & 297  (28)   & 558  (59)   & 296  (29)   & 571  (64)   & 290  (27)   & 555  (59)   \\\hline
\multicolumn{1}{|c}{$\mathbf{\omega_{01}^2}$} & \multicolumn{2}{|c|}{0.044 (0.012)} & \multicolumn{2}{|c|}{0.040 (0.011)} & \multicolumn{2}{|c|}{0.044 (0.013)} \\\hline
\multicolumn{1}{|c}{$\mathbf{\omega_{12}^2}$} & \multicolumn{2}{|c|}{0.083 (0.034)} & \multicolumn{2}{|c|}{0.015 (0.023)} & \multicolumn{2}{|c|}{0.041 (0.034)} \\\hline
\multicolumn{1}{|c}{$\mathbf{\omega_{13}^2}$} & \multicolumn{2}{|c|}{0.258 (0.089)} & \multicolumn{2}{|c|}{0.199 (0.098)} & \multicolumn{2}{|c|}{0.073 (0.059)} \\\hline
\multicolumn{1}{|c}{$\mathbf{\omega_{31}^2}$} & \multicolumn{2}{|c|}{0.044 (0.040)} & \multicolumn{2}{|c|}{0.024 (0.031)} & \multicolumn{2}{|c|}{0.079 (0.043)} \\\hline
\multicolumn{1}{|c}{$\mathbf{\omega_{43}^2}$} & \multicolumn{2}{|c|}{0.296 (0.095)} & \multicolumn{2}{|c|}{0.128 (0.071)} & \multicolumn{2}{|c|}{0.203 (0.068)} \\\hline
\multicolumn{1}{|c}{$\mathbf{\omega_{Q_1}^2}$} & \multicolumn{2}{|c|}{0.161 (0.040)} & \multicolumn{2}{|c|}{0.173 (0.043)} & \multicolumn{2}{|c|}{0.158 (0.041)} \\\hline
\multicolumn{1}{|c}{$\mathbf{S}$ } & \multicolumn{2}{|c|}{$4.34\cdot10^{-3}$ ($0.38\cdot10^{-3}$)} & \multicolumn{2}{|c|}{$4.89\cdot10^{-3}$ ($0.74\cdot10^{-3}$)} & \multicolumn{2}{|c|}{$6.98\cdot10^{-4}$ ($2.61\cdot 10^{-4}$)} \\\hline
\multicolumn{1}{|c}{$\mathbf{\alpha}$ } & \multicolumn{2}{|c|}{---} & \multicolumn{2}{|c|}{---} & \multicolumn{2}{|c|}{3.09 (0.95)} \\\hline
\multicolumn{1}{|c}{$\mathbf{\sigma}$ } & \multicolumn{2}{|c|}{---} & \multicolumn{2}{|c|}{---} & \multicolumn{2}{|c|}{0.222 (0.027)} \\\hline
\end{tabular}
\end{footnotesize}
\caption{{\footnotesize Estimated parameters (and standard errors of the estimates) for the NLME model explained in equations~\mbox{\eqref{eq:k01i}-\eqref{eq:Q1i}}. In the left panel the parameters of the ODE model~\mbox{\eqref{eq:stateNLME_ODE}-\eqref{eq:OutputWithNoise}} are presented when all experimental data were used. In the middle the estimated parameters are presented when the samples at $4,\ 8,\ 10,\ 15$, and $45$ minutes and $3,\ 6$, and $8$ hours were removed. The right panel consists of the corresponding estimated parameters, together with the parameters $\alpha$ and $\sigma$, when the SDE model~\mbox{\eqref{eq:Tracee_SDE}-\eqref{eq:OutputWithNoise_SDE}} was applied to all data.}}
\label{table:results_NLME_expData}
\end{center}
\end{table}

In addition to estimating the parameters using all experimental data, we also estimated the parameters using a smaller set of data, in which the data points at $4,\ 8,\ 10,\ 15$, and $45$ minutes and $3,\ 6$, and $8$ hours were removed. Thus, at most eight samples per individual were used for the smaller set. The reason for estimating the parameters for a smaller data set was to test if population parameters can be reliably estimated even if the number of samples in the time series is low for each separate individual. It should be mentioned that the two stage approach was also applied to the smaller data set. In that case the spread of the estimated parameters was much larger between individuals compared to when all data were used and at least one parameter reached the predefined lower- or upper bounds for most individuals.

\begin{figure}[b!]
\begin{center}
\includegraphics[width=\textwidth,height=12cm]{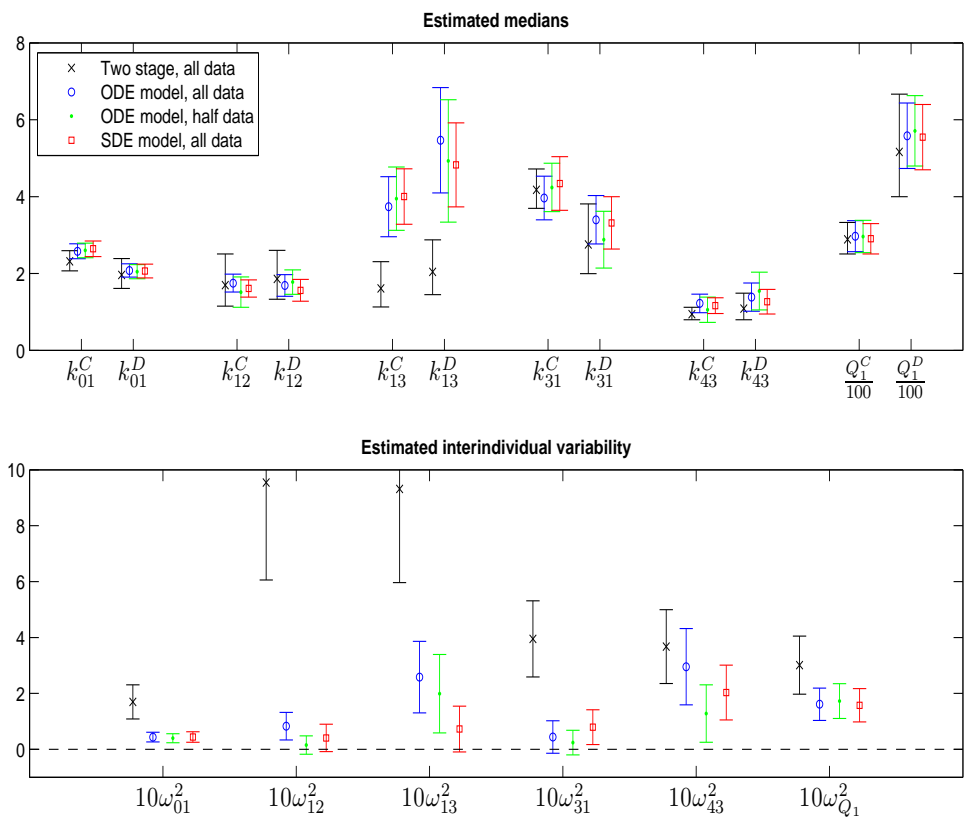}
\end{center}
\caption{{\footnotesize $85\%$ confidence intervals of the parameters that describe the medians of the estimated parameters and the interindividual variations for the three groups, respectively. Medians denoted by 'x' represent the two stage approach (upper values for the intervals of $10\omega_{12}^2$ and $10\omega_{13}^2$ are $13.0$ and $12.7$, respectively). The 'o':s represent the ODE based NLME approach applied to all data, the '$\cdot$':s represent the ODE based NLME approach applied to the smaller set of data, and the }'\tiny{$\square$}{\footnotesize ':s represent the SDE based NLME model applied to all data.}}
\label{fig:CIs}
\end{figure}

In Table~\ref{table:results_NLME_expData} and Figure~\ref{fig:CIs} it can be seen that using at most eight samples per individual gave no disjoint $85\%$ confidence intervals compared to when all data were used (we used the standard errors of the estimated parameters to obtain confidence intervals for the NLME model). Most parameters are also very similar in general, but the estimates of $\omega_{12}^2,\ \omega_{31}^2$, and $\omega_{43}^2$ deviates rather much for the two data sets and the overlapping confidence intervals are due to large uncertainties of the estimates. The estimated standard errors of the average parameters and the intraindividual noise parameter, $S$, were in general higher when only half the data set was used, indicating more uncertain parameter estimates.

We were also interested in comparing the estimated population parameters from the two stage approach with the results from the NLME approach when all data were used. As can be seen in Figure~\ref{fig:CIs} some estimated values differ between the two approaches, and the estimated medians $k_{13}^C,$ and $k_{13}^D$  have disjoint $85\%$ confidence intervals. Moreover, all the interindividual variances are estimated to be smaller for the NLME model and $\omega_{01}^2,\ \omega_{12}^2,\ \omega_{13}^2,$ and $\omega_{01}^2$ are smaller with statistical significance.

Observe that the estimated population parameters for the two stage approach in Section~\ref{Sec:leucine model - parEst} were trimmed (by removing estimates of parameters significantly different from most other individuals) before computing population averages and variations of the parameters. If all estimated parameters are accounted for in that case the interindividual differences are even larger. Then the NLME approach predicts a significantly smaller interindividual variation also for $k_{43}$.

With the two stage approach we discovered that $Q_1$ differed between the groups. For the ODE based NLME model we see that there are significant differences for $k_{01}$ and $Q_1$.

It is worth mentioning that the confidence intervals for the population parameters of the lognormal distributions in the NLME model may be calculated as in Section~\ref{Sec:leucine model - parEst}, by taking every individual into account. Doing so, the $85\%$ intervals were in general more narrow compared to using the ``standard error approach'' and, in addition to $k_{01}$ and $Q_{1}$, also $k_{13}$ and $k_{31}$ have disjoint confidence intervals between the groups. Thus, when confidence intervals of the medians of the population parameters are computed by taking every individual into account, we conclude that both the leucine levels in the plasma and leucine kinetics differ between control individuals and individuals suffering from diabetes type~$2$.

\subsection{The SDE based NLME model} \label{Sec:leucine model - popSDE}
We extend the ODE model by assuming that the tracee in compartment~$1$ is fluctuating around a steady state value. The state variable $Q^i_{1,t}$ is modelled with a mean-reverting Ornstein-Uhlenbeck process as discussed in Section~\ref{Sec:Theory-NLME_SDE} for a one-state compartmental model. The reason for testing the model under this assumption is that the amount of amino acid material is probably not exactly constant in the body, but varies over time. Daily rhythms of leucine levels have also earlier been reported (see, e.g., \citealt{Lavie:2004}).

The SDE model takes the form
\begin{align}
\mathrm{d}Q_{1,t}^i & = \alpha(Q_{1,0}^i-Q_{1,t}^i)\mathrm{d}t + \sigma Q_{1,0}^i\mathrm{d}W_t, & Q_1^i(0) & \sim N\left(Q_{1,0}^i,\frac{(\sigma Q_{1,0}^i)^2}{2\alpha}\right), \label{eq:Tracee_SDE} \\
\mathrm{d}q^i_t & = (K^iq^i_t)\mathrm{d}t, & q^i(0) & =[q_{1,0}^i,0,0,0]^T, \label{eq:stateNLME_SDE} \\
Y^i_k & = \log(\frac{q^i_{1,k}}{Q^i_{1,k}}) + v_k^i, & v_k^i & \sim N(0,S), \label{eq:OutputWithNoise_SDE}
\end{align}
where $Q_{1,0}^i$ corresponds to $Q_1^i$ in Section~\ref{Sec:leucine model - pop}. Thus, compared to the ODE model, two new parameters, $\alpha$ and $\sigma$, must be estimated.

\subsubsection{Estimated parameters} \label{Sec:leucine model - popSDE - parEst}
When using the smaller set of data, the estimate of $\sigma$ converged to zero, and the SDE model was reduced to the ODE model. Therefore we only present the parameter estimates when the model was used on the complete data set.

Overall, the SDE model produced similar parameter estimates as the ODE model (see Table~\ref{table:results_NLME_expData} and Figure~\ref{fig:CIs}). However, $\omega_{12}^2,\ \omega_{13}^2,\ \omega_{31}^2$, and $S$ deviate quite much, but only $S$ have overlapping confidence intervals. Interestingly, the parameters that describe interindividual variability are, in general, smaller than for the ODE model.

Like the ODE based NLME model, the SDE model predicts group differences for $k_{01}$ and $Q_1$ when the ``standard error'' approach was used to compute confidence intervals and for $k_{01}$, $k_{13}$, $k_{31}$, and $Q_1$ when all individual estimates were used.

We observed strong correlations between the parameters describing the intraindividual variations of the model. The estimated correlations were $0.79$ between $\alpha$ and $\sigma$, $-0.40$ between $\alpha$ and $S$, and $-0.69$ between $\sigma$ and $S$.

\subsection{Comparison between ODE and SDE results} \label{Sec:ODEvsSDE}
\begin{figure}[b!]
\noindent\makebox[\textwidth]{ 
\includegraphics[width=1.1\textwidth,height=7cm]{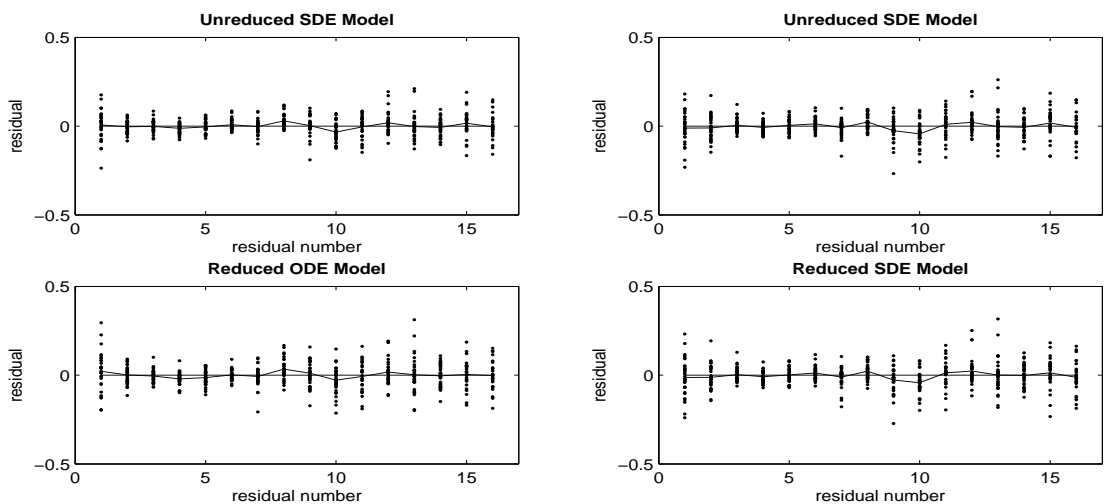}
}
\caption{{\footnotesize The residuals for all $34$ individuals at the sixteen sampled time points. The lines correspond to mean values. In the lower plots the reduced models described in Section~\ref{Sec:reducedModels} are used.}}
\label{fig:residuals_subplot}
\end{figure}

Both the ODE- and SDE based NLME models gave good fits to data (see the upper plots in Figure~\ref{fig:residuals_subplot}) and small correlations between residuals (when SDEs are used the term residual refers to the observed data point subtracted by the one step prediction obtained from the extended Kalman filter at the given time point). The difference between the two approaches is that the SDE model predicts a varying leucine level. Let $z_t^i=Q_{1,t}^i/Q_{1,0}^i$ be the normalized process described by $\mathrm{d}z_t^i=\alpha(1-z_t^i)\mathrm{d}t+\sigma\mathrm{d}W_t$. Then $z_t^i$ has variance $\sigma^2/(2\alpha)$, thus the leucine level is predicted to vary with a standard deviation of $100\times\sigma/\sqrt{2\alpha}$ percent of the predicted average value. With the estimated values of $\sigma$ and $\alpha$ we get a standard deviation of $0.089$.

\begin{figure}[b!]
\begin{center}
\includegraphics[width=14cm,height=6cm]{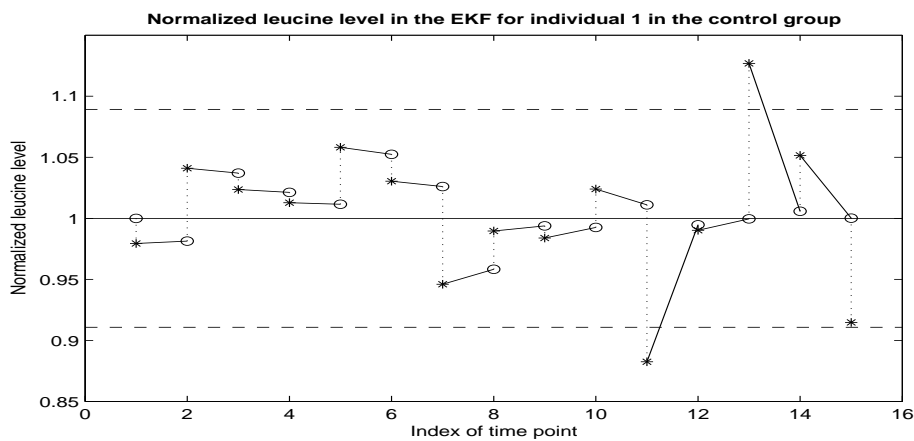}
\end{center}
\caption{{\footnotesize The normalized free leucine level in the EKF for an individual in the control group. The dashed lines indicate $1 \pm 0.089$, i.e., the mean plus/minus the standard deviation of the process at each time point. For this individual there is a missing data point at $8$ hours, i.e., the data are taken at $2,\ 4,\ 6,\ 8,\ 10,\ 12,\ 15,\ 20,\ 30$, and $45$ minutes and $1,\ 2,\ 3,\ 4$, and $6$ hours. Observe that when the interval between measurements increase, the jumps away from the mean becomes longer, indicating an increase in the uncertainty of the state.}}
\label{fig:EKF}
\end{figure}

In Figure~\ref{fig:EKF} we visualize how the level of free leucine in the body changes in the extended Kalman filter for a single individual. First the model predicts the level to be at the mean level. Then an update is performed at the first measurement time point and the model makes a new prediction at the next time point, given the previous update. The procedure is then iterated until the last time point is reached. Observe that we at the x-axis have the index of each time point, not the actual times. In the plot the leucine level is normalized as described above.

In the output functions,~\eqref{eq:OutputWithNoise} and~\eqref{eq:OutputWithNoise_SDE}, it is assumed that the errors $v_k^i$ are Gaussian and independent. The first assumption has been tested by applying the Lilliefors test function \emph{lillietest} in MATLAB to the residuals. With significance level~$0.05$ an assumption about normality can be rejected for four individuals for the ODE model and one individual for the SDE model (in these individuals we observed one or two residuals significantly different from the other, indicating outlying observations).

One way to investigate the independence assumption is by means of the autocorrelations between the residuals (one step prediction errors). That was also done in \cite{Moller:2010} and \cite{Overgaard:2007}, where they concluded that the residual correlations decreased for their proposed SDE models.

However, the autocorrelation of the residuals does not give any information about correlations between specific sample points, but is more a general representation of the correlations for specific lags. When analyzing multiple individuals sampled at the same times we can instead compute the sample correlations of the residuals for specific time points (the correlation between residuals at time $t_j$ and $t_k$ is calculated as~\mbox{$r_{jk}=\sum_{i=1}^N\frac{(\epsilon_j^i-\bar{\epsilon}_j)(\epsilon_k^i-\bar{\epsilon}_k)}{(N-1)s_js_k}$}, where $\bar{\epsilon}_k$ is the mean of all the residuals at time $t_k$, and $s_k$ is the sample standard deviation). This gives a more detailed representation of the residual correlation structure. For the NLME models presented in Sections~\ref{Sec:leucine model - pop} and~\ref{Sec:leucine model - popSDE} the correlations between the residuals are small. A correlation matrix for the residuals from the ODE model is presented in the left hand plot in Figure~\ref{fig:internalCorrelations}.

\subsubsection{Reduced ODE- and SDE based models} \label{Sec:reducedModels}
If an ODE based model is less accurate it may produce subsequent over- or underpredictions of the data, thus producing correlated residuals. A method to reduce these correlations may be to introduce system noise by means of SDEs in the model. Since the ODE model described above seems to produce good fits to data we have investigated this for a reduced version of the leucine model where $\omega_{12}$ and $\omega_{13}$ were fixed to zero (the reason to choose these parameters is that a sensitivity analysis indicate that $k_{12}$ and $k_{13}$ have the least influence on the objective function). We call the new models the \emph{reduced} ODE- and SDE models, respectively.

\begin{figure}[b!]
\begin{center}
\includegraphics[width=\textwidth,height=6cm]{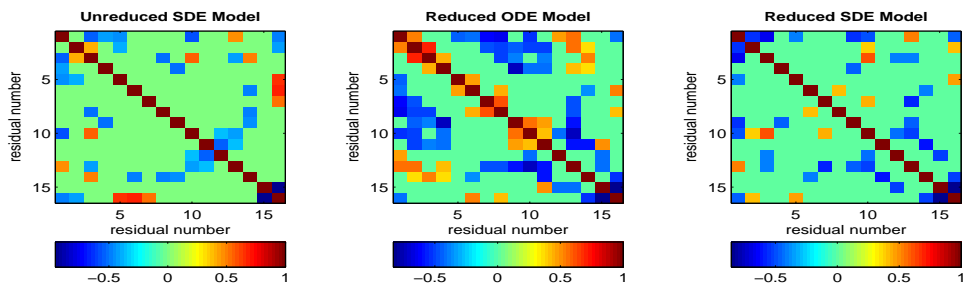}
\end{center}
\caption{{\footnotesize Correlation matrices for the residuals for the ODE based NLME model where all parameters vary between individuals, and the reduced ODE- and SDE models respectively. Only the correlations significantly different from zero ($p<0.05$) are highlighted and correlations with $p>0.05$ are set to zero. The p-values are calculated by testing the hypothesis that the correlation $r_{jk}$ is zero against the alternative $r_{jk}\neq 0$ as explained in \cite{Larsen:2001}.}}
\label{fig:internalCorrelations}
\end{figure}

The sample correlation matrices for the residuals of the reduced models are also presented in Figure~\ref{fig:internalCorrelations}. We observe that the residuals are less correlated for the SDE model. The most striking difference between the two models is that the correlations between adjacent residuals are in general positive for the ODE model but not for the SDE model. The conclusion to be drawn from this is that the SDE model well captures the involved mechanisms and that the remaining noise and uncertainty can be described well by normally distributed random variables and Wiener processes.

\section{Discussion}
In this article population models for leucine kinetics in blood plasma have been proposed. The models are based on differential equations to account for the dynamic behaviour of leucine molecules. The first models are based on ordinary differential equations where the tracee material is assumed to be in a steady state. The NLME model is then extended so that the amount of tracee material is modelled by an Ornstein-Uhlenbeck process to account for fluctuating tracee material.

The estimated population average parameters did in general not change much when SDEs were used instead of ODEs in the NLME settings, but the interindividual variabilities were in general estimated to be smaller for the SDE model. In systems where the tracee level fluctuates much around a steady state during the time of the experiments, we believe that errors in the estimated physiological parameters may occur if the fluctuations are not accounted for in the model. We have performed simulation studies for the one-compartment SDE model~\mbox{\eqref{eq:TraceeOneCompModel}-\eqref{eq:Q0iOneComp}} that supports this hypothesis. Especially the parameters that describe the interindividual variability tend to be overestimated when an ODE based model is used for data generated from the SDE model.

For the constrained model ($\omega_{12}=\omega_{13}=0$) the residuals were correlated when ODEs were used. Correlations between the residuals indicate that the assumption that the errors $v_k$ and $v_l$ are independent when $k\neq l$ is incorrect. For the constrained SDE model the correlations were smaller. This explains how SDE may be used as a tool to handle non-explicitly modelled effects in the system. For the present leucine kinetics model there is no obvious reason not to use the unconstrained model, but it should be mentioned that this model requires two more parameters to be estimated for each individual and the computational time needed to estimate the parameters with the unconstrained model increased significantly. For larger models it may be computationally too demanding to estimate all the parameters for each individual.

Other SDE models may be used to describe the system. One approach, that was recently used by \cite{Moller:2010}, is to add a zero-mean Ornstein-Uhlenbeck process to the model output to allow for a more flexible model output structure. We have tested that approach and the results indicate that such a model may equally well be used to describe the system. However, that model does not have a direct physiological description and we decided to only present the results for the variable tracee model.

The parameters of the ODE based NLME model were also estimated using half the data for each individual, and the results were similar to the ones obtained using all data. This indicates that the number of samples needed per individual does not necessarily have to be large for the ODE based NLME model. This may be of importance in experimental planning since taking many repeated measurements may be expensive or troublesome. The SDE model could not be applied when the smaller data set was used, since $\sigma$ converged to zero in that case, resulting in the ODE model. Thus, when the EKF is applied to the SDE based leucine kinetics model, the number of data points per individuals needs to be sufficiently large. \cite{Overgaard:2005} show that, when the number of individuals is large, only three samples per individual are enough to estimate parameters for a simple one-compartment SDE model. Further studies are needed to investigate the number of individuals and data points per individual that is needed to estimate the parameters for our model. Of particular interest is to investigate the model performance for only a few data points per individual. To capture the dynamical behaviour, it is then important that the measurements are not taken at the same time points for each individual.

Differences were observed when the estimated parameters from the two stage approach were compared to the results from the NLME models. Population median estimates differed for some parameters but especially the interindividual variations were estimated to be much smaller for the NLME models. It has earlier been observed that, for some models, the two stage approach produce biased average parameters (see, e.g., the work by \citealt{Sheiner&Beal:1983}). That the interindividual variation is overestimated with the two stage approach has earlier been observed in a large number of publications, e.g., in \cite{Olofsen:2004,Sheiner&Beal:1980,Sheiner&Beal:1983,Steimer:1984}. One explanation why the two stage approach seems to produce biased estimates of population parameters is that the variance of the average parameters depend both on intra- and interindividual variations. This is shown in \cite{Olofsen:2004} for a simple linear model. NLME models separate these two kinds of variations.

We have investigated the model parameters for two groups of individuals, one group of diabetic individuals and one control group. The estimated average behaviour of the parameters differs between the groups, and especially $Q_1$ is much larger for the diabetic group. Larger levels of free leucine for diabetic individuals have been reported earlier and the amount of leucine is associated with the level of insulin treatment and magnitude of hyperglycemia for the individuals~\citep{Gougeon:2008}. Moreover leucine degradation and protein uptake of free leucine ($k_{01}$ and $k_{31}$) is smaller for the diabetic group and leucine production due to increased proteolysis is larger. These findings are also in agreement to what has been reported earlier~\citep{Gougeon:2008}.

There are also other physiological differences between the groups (age, body weight, level of obesity, etc.) that may influence the parameters and a thorough investigation how such covariates associates with model parameters is a topic of further research. Such knowledge is important when NLME results are used in predictive studies and when parameters are estimated for single individuals, especially when data are very sparse. The NLME framework can then be used by simply maximizing the individual loglikelihood~\eqref{Eq:indLikelihood} with all the population parameters fixed to results from earlier population studies. Note that predictions of individual parameters can then be performed for very few data points (even a single data point) but the predictive performance obviously increase for individuals that have been densely sampled.

To analyze the identifiability of the model parameters we have performed studies where data were simulated for a given set of parameters. Given the simulated data we estimated the parameters and they were in general similar to the true parameters, except that $\alpha$ and $\sigma$ were somewhat overestimated.

It would be interesting to apply other filters, which are adapted to nonlinear models. However, in this particular model it is only the output function that is nonlinear, and we believe that the EKF is a good choice.

The use of nonlinear mixed effects models were originally used in the area of pharmacokinetics and pharmacodynamics. Recently the number of applications has been growing and the present paper is a first step towards larger NLME models in the metabolic arena, focusing on studying lipoprotein kinetics. Using SDEs may be a good way to account for model approximations and/or true variations of model state variables and parameters.

\section{Acknowledgments}
The authors are thankful to Dr Marita Olsson (Chalmers University of Technology) for discussions concerning some of the statistical contents of this paper and Professor Marja-Riitta Taskinen (University of Helsinki) for providing data.

\section{Funding}
This research is funded by the Swedish Foundation for Strategic Research both directly and via the Gothenburg Mathematical Modelling Centre and the Sahlgrenska Center for Cardiovascular and Metabolic Research. It has also been supported by grants from the European Commission 6th Framework Programme
(BIOSIM, grant No 005137) and 7th Framework Programme (UNICELLSYS, grant No 201142) as well as from grants from Novo-Nordisk foundation.


{\footnotesize}{\small}
\bibliographystyle{apalike}          
\bibliography{references}            


\end{document}